\pdfoutput=0
\documentclass[11pt,a4paper]{article}
\usepackage{jcappub}
\usepackage{bbm,bm}
\usepackage{xcolor}
\usepackage{cancel}
\usepackage{empheq}
\usepackage{scalerel}
\usepackage{verbatim}

 \allowdisplaybreaks[1]

\newcommand{\la}[1]{\label{#1}}
\newcommand{\be}{\begin{equation}}
\newcommand{\ee}{\end{equation}}
\newcommand{\ba}{\begin{eqnarray}}
\newcommand{\ea}{\end{eqnarray}}
\newcommand{\bi}{\begin{itemize}}
\newcommand{\ei}{\end{itemize}}
\newcommand{\nn}{\nonumber \\}

\newcommand{\eq}{eq.~}
\newcommand{\eqs}{eqs.~}

\newcommand{\se}{sec.~}

\newcommand{\nr}[1]{(\ref{#1})}

\newcommand{\rmi}[1]{{\mbox{\scriptsize #1}}}
\newcommand{\rmii}[1]{{\mbox{\tiny\rm{#1}}}}

\renewcommand{\vec}[1]{{\bf #1}}
\newcommand{\mpl}{m_\rmii{pl}} 

\def\lsi{\raise0.3ex\hbox{$<$\kern-0.75em\raise-1.1ex\hbox{$\sim$}}}
\def\gsi{\raise0.3ex\hbox{$>$\kern-0.75em\raise-1.1ex\hbox{$\sim$}}}

\newcommand{\Tint}[1]{{\hbox{$\sum$}\!\!\!\!\!\!\!\int\,}_{\!\!\!\!\raise-0.9ex\hbox{$\scriptstyle{#1}$}}}
\newcommand{\Tinti}[1]{{{\Sigma}\!\!\!\!\raise0.3ex\hbox{$\int$}_\rmii{${#1}$}}}
\newcommand{\hide}[1]{ }

\newcommand{\deltabar}{\raise-0.02em\hbox{$\bar{}$}\hspace*{-0.8mm}{\delta}}
\newcommand{\ddeltabar}{\raise-0.18em\hbox{$\bar{}$}\hspace*{-0.8mm}{\delta}}
\renewcommand{\H}{\mathcal{H}}

\newcommand{\Q}{\mathcal{Q}}

\newcommand{\R}{\mathcal{R}}

\newcommand{\X}{\mathcal{X}}
\newcommand{\Y}{\mathcal{Y}}

\newcommand{\G}{\mathcal{G}}

\newcommand{\der}{,}

\newcommand{\barpPi}{\Pi}

\newcommand{\rmO}{{\mathcal{O}}}
\newcommand{\ord}{\mathcal{O}}
 
\newcommand{\now}{\rmi{0}}

\newcommand{\tensor}{\rmi{t}} 

\newcommand{\rev}[1]{#1}

\makeatletter \@addtoreset{equation}{section} \makeatother
\renewcommand{\theequation}{\arabic{section}.\arabic{equation}}
\makeatletter
\renewcommand\section{\@startsection {section}{1}{\z@}%
                                   {-5.5ex \@plus -1ex \@minus -.2ex}
                                   {2.3ex \@plus.2ex}%
                                   {\normalfont\large\bfseries}}
\renewcommand\subsection{\@startsection{subsection}{2}{\z@}%
                                     {-3.25ex\@plus -1ex \@minus -.2ex}%
                                     {1.5ex \@plus .2ex}%
                                     {\normalfont\normalsize\bfseries}}
\renewcommand\thesection {\@arabic\c@section}
\renewcommand\thesubsection   {\thesection.\@arabic\c@subsection}
\renewcommand{\@seccntformat}[1]{%
\csname the#1\endcsname.\hspace{1.0em}}
\makeatother

\begin{document}

\flushbottom


\begin{flushright}
May 2026
\end{flushright}

\vspace*{-0.9cm}

\arxivnumber{2512.04482}

\title{\boldmath \Large
  General SIGW source for reheating dynamics
}


\author[a]{\large M.~Laine,}
\author[b,1]{\large S.~Procacci\hspace*{0.3mm}}%
\note{%
 Previously at: 
 D\'epartement de Physique Th\'eorique, 
 Universit\'e de Gen\`eve, 
 24 quai Ernest Ansermet, 
 CH-1211 Gen\`eve 4, Switzerland
}


\affiliation[a]{
AEC, 
Institute for Theoretical Physics, 
University of Bern, \\ 
Sidlerstrasse 5, CH-3012 Bern, Switzerland
}


\affiliation[b]{
\hspace*{0.3mm}KKM BKW AG, M\"uhleberg, Switzerland
}

\emailAdd{laine@itp.unibe.ch}
\emailAdd{simona.procacci@pm.me}


 

 
%
\abstract{
Working in an arbitrary gauge, we derive 
the source term for scalar-induced gravitational waves (SIGW) 
valid during a general reheating epoch. 
The dominant energy component is allowed to transition smoothly from 
an inflaton field to a set of fluids, 
possibly via a period of matter domination. 
Gauge invariance is verified up to second order. 
}
%



 


\maketitle

%
\section{Introduction}

The observed anisotropies in 
the cosmic microwave background (CMB) and large-scale structure (LSS) 
suggest that the early universe featured a nearly scale-invariant
spectrum of density (``scalar'') perturbations. 
However, CMB and LSS observations 
constrain the amplitude of the spectrum only 
in a relatively narrow range of small momenta,
leaving over the possibility of 
large deviations from scale invariance 
at larger momenta. The idea of having large-momentum peaks 
is exciting, because they might facilitate the formation of  
exotic structures, like early galaxies or primordial black holes,
both actively searched for. 

At second order in perturbation theory, 
peaked features in scalar power spectra 
(whether of curvature or isocurvature type)
lead also to the enhanced generation of 
gravitational waves~(cf.~refs.~\cite{kt,rai} for key notions, 
refs.~\cite{mollerach,matarrese,nakamura,wands,sigw} 
for influential early works, 
and ref.~\cite{domenech} for a review). 
This in turn opens a door for their experimental search 
(cf., e.g., refs.~\cite{bian,pieroni,lisa,lvg}). 
This mechanism is conventionally referred to as 
``scalar-induced gravitational waves'' (SIGW). 

The theoretical 
computation of the SIGW spectrum consists of two parts. The first is 
the derivation of the source term, appearing on the right-hand
side of the inhomogeneous differential
equation satisfied by gravitational perturbations. The second is
the solution of the differential equation with 
the Green's functions method, 
and the evaluation of the gravitational-wave energy density 
at the time of observation  
(cf., e.g., refs.~\cite{racco,terada,inomata1,inomata2,gong,gd2}).
Of particular intrigue are gravitational waves sourced 
at relatively late times, 
for instance during a period of matter domination at the end of inflation 
(cf., e.g., refs.~\cite{md1,md15,md2,white,chicago,chluba,iso2,polter}). 
In the present paper, 
we concentrate on the first part, 
the derivation of the source term, however one application
that we have in mind is indeed related to the reheating epoch
(cf.\ \se\ref{se:concl}). Our source term is general
enough to also have other 
applications, such as gravitational waves generated by 
irrotational velocity perturbations 
($\delta\vec{v} = - \nabla \delta v$ so that 
$\nabla\times \delta\vec{v} = \vec{0}$)
during a first-order phase transition in the radiation-dominated epoch. 

In every SIGW-related computation, 
gauge dependence is an important issue. 
When deriving the source term, the early 
SIGW papers had fixed a gauge. 
Subsequently, relaxing the assumption, the result
inevitably appears to depend on the gauge
(cf.,\ e.g.,\ refs.~\cite{gauge,Tomikawa:2019tvi,Sipp:2022kmb,%
Ali:2023moi,Yuan:2024qfz}). This then 
questions the physical significance of the SIGW mechanism. 

However, an apparent 
gauge dependence is not the end of the story. The fields
that appear in the equations are not independent of each other, 
but they are interrelated by the Einstein equations. It could be that when 
posing a well-defined initial-value problem, solving the evolution
equations, and evaluating a consistently defined gravitational-wave
energy density at the end, the apparent gauge dependence drops out. 
This has led to a large body of work, 
ranging from theoretical considerations about how to properly
define the observable, 
via arguments about the superiority of certain ``well-behaved'' gauges,  
to practical checks, 
suggesting that in the end the SIGW signal
should be physical after all
(cf.,\ e.g.,\ refs.~\cite{Domenech:2017ems,DeLuca:2019ufz,Inomata:2019yww,%
Yuan:2019fwv,Lu:2020diy,Chang:2020tji,%
Ali:2020sfw,Chang:2020iji,counterterm1,%
Domenech:2020xin,Cai:2021ndu,%
Ota:2021fdv,counterterm2,Yuan:2025seu,Kugarajh:2025pjl,%
Ali:2025xvj,Domenech:2025ccu}). 
A clearly expressed summary on many of these approaches
(and their possible shortcomings) 
can be found in ref.~\cite{counterterm2}.

Despite the work done, 
there are arguably gaps to fill. 
One of them is that in the literature, 
the source term is 
regularly represented by a single curvature perturbation. However, at
horizon re-entry, as they start to undergo acoustic oscillation, 
different scalar degrees of freedom 
(inflaton field, fluid velocities, 
fluid temperatures)
obtain independent dynamics, 
even in a minimal system in which no primordial isocurvature
perturbations are present. 
As we will show, 
this changes the structure of the source term (cf.\ \eq\nr{sigw_2}). 

The concrete goal of the present study is 
to enrich the gauge-independent SIGW setup so that a general
reheating epoch can be investigated. Specifically, 
we keep track of four scalar degrees
of freedom in the metric, an energy-momentum tensor containing both
an inflaton field and fluids, and 
an expanding background whose nature is left unspecified. 
Following ref.~\cite{counterterm2}, 
we show that the resulting equation is gauge invariant 
at second order. However, it does contain non-propagating
metric perturbations evaluated 
at the observation time, whose values are ambiguous. 
We subsequently argue that
unambiguous results can be obtained
by carrying out measurements 
in a locally flat patch, though the precise definition
of an observable is not addressed in this work. 

Our presentation is organized as follows. 
After defining our setup (cf.\ \se\ref{ss:setup}), 
we work out the left-hand side of the Einstein equations, 
i.e.\ the Einstein tensor (cf.\ \se\ref{ss:lhs}). 
Then we turn to the right-hand side of the Einstein equations, 
i.e.\ the energy-momentum tensor (cf.\ \se\ref{ss:rhs}). 
This requires more thought 
than the left-hand side, as background and first-order
relations need to be properly inserted in order to obtain
a transparent outcome. The key part is to 
equate the two sides, 
finding a substantial cancellation 
between gauge-dependent terms (cf.\ \se\ref{ss:equate}).
Subsequently, we show that the non-cancelled remainder 
is gauge-invariant (cf.\ \se\ref{se:interp}), 
\rev{and compare
its right-hand side with previous results in the literature
(notably, refs.~\cite{sigw,chicago}).} 
We conclude this study with 
a summary and outlook (cf.\ \se\ref{se:concl}).

%
\section{Derivation of the effective source term}
\la{se:derivation}

%
\subsection{General setup}
\la{ss:setup}

Our goal is to derive a SIGW source term that is broadly applicable 
to a single-field inflationary scenario. 
Notably, we allow for a smooth transition from the inflationary epoch
to a heating-up period, as well as to  
the subsequent radiation or matter-dominated expansion. To this aim, we
consider an energy-momentum tensor composed of a scalar field, $\varphi$,
and a set of fluids, numbered by $n \in \{1,...,n^{ }_{\rm tot}\}$, 
with their respective 
flow velocities, $u^{ }_{n \mu}$. 
The overall energy density, $e$, and pressure, $p$, include
contributions 
from the energy densities and pressures of 
the various fluids,  
and 
from the self-interaction potential of $\varphi$, 
$e \equiv \sum^{n^{ }_{\rm tot}}_{n = 1} e_n + V$, 
$p \equiv \sum^{n^{ }_{\rm tot}}_{n = 1} p_n - V$. 
It is convenient to define $e^{ }_0 \equiv V$, 
$p^{ }_0 \equiv -V$, $u^{ }_0 \equiv (1,\vec{0})$, 
and $\sum_n \equiv \sum^{n^{ }_{\rm tot}}_{n = 0}$.
We also incorporate an
anisotropic stress, $\Pi^{ }_{\mu\nu}$, which could involve 
shear-viscous corrections to fluid motions; for the moment we keep it 
as such, and return {\em a posteriori} to its interpretation
(cf.\ \se\ref{se:interp}).
With these ingredients, the energy-momentum tensor takes the form 
\be
 T^{ }_{\mu\nu}
 \; 
  \underset{\rmii{ }}{
  \overset{\rmii{ }}{\equiv}} 
 \; 
 \varphi^{ }_{,\mu}\varphi^{ }_{,\nu}
 - \frac{ g^{ }_{\mu\nu}\, 
 \varphi^{ }_{,\alpha} \varphi^{,\alpha}_{ } }{2} 
 + \sum_n ( e^{ }_{n} + p^{ }_{n} ) 
   \,  u^{ }_{n \mu} u^{ }_{n \nu}
 + \sum_n p^{ }_{n} \, g^{ }_{\mu\nu}
 + a^2_{ } \Pi^{ }_{\mu\nu}
 \;, \la{Tmunu_mixed}
\ee
where 
$
 (...)^{ }_{\der \mu} \equiv \partial^{ }_{\mu} (...)
$; 
the metric, $g^{ }_{\mu\nu}$, 
is assumed to have the signature ($-$$+$$+$$+$); 
and $a$ is the cosmological scale factor. 
We note that the 
kinetic energy of $\varphi$ is kept apart, as it appears in 
a different covariant structure than the fluid terms. 

The metric that appears in \eq\nr{Tmunu_mixed}
can be parametrized as 
\be
  g^{ }_{\mu\nu} \;\equiv\; a^2_{ }(\tau) \begin{pmatrix}
 -1-2h^{ }_0 & -h^{ }_j \\
 -h^{ }_i    & (1-2h^{ }_\rmii{D})\,\delta^{ }_{ij}+2\vartheta^{ }_{ij}
 \end{pmatrix} 
 \ , \la{g_munu}
\ee
where $\tau$ denotes the conformal time coordinate,
and the function $ \vartheta^{ }_{ij} $ is defined to be traceless. 
Up to this point, no approximation has been made. From the metric, 
we can derive the Einstein tensor ($G^{ }_{\mu\nu}$), and then impose
the Einstein equations between $G^{ }_{\mu\nu}$ and 
$T^{ }_{\mu\nu}$, 
\be
 G^{ }_{\mu\nu} \; = \; 8 \pi G \, T^{ }_{\mu\nu}
 \;, \la{einstein}
\ee
where $G = 1/\mpl^2$ is the Newton constant and 
$\mpl^{ } = 1.2209 \times 10^{19}_{ }\,\mbox{GeV}$ is the Planck mass. 
We place both indices down
in $G^{ }_{\mu\nu}$ and $T^{ }_{\mu\nu}$, as this guarantees
that the tensors are symmetric. 

%
\subsection{Left-hand side of the Einstein equations}
\la{ss:lhs}

We expand the left-hand side of 
\eq\nr{einstein} in a perturbative approach.  
In considerations of SIGW, it is sufficient 
to restrict to scalar and tensor fluctuations
of the metric, with ``scalar'' and ``tensor'' 
referring to transformations in a plane
transverse to $\partial^{ }_i$. Concretely, we write
\be
 - h^{ }_i
 \; \equiv \;
 h^{ }_{\der i}
 \;, \quad
 \vartheta^{ }_{ij}
 \; \equiv \;
 \vartheta^\tensor_{ij}
 + 
 \vartheta^{ }_{\der ij}
 - \frac{\delta^{ }_{ij}}{3}\, \vartheta^{ }_{\der kk} 
 \;, \quad
 Y 
 \; \equiv \;  
 h^{ }_\rmii{D} + \frac{\vartheta^{ }_{\der kk}}{3} 
 \;, \la{scalar}
\ee
where latin indices are assumed spatial; 
the tensor part
satisfies $ \vartheta^\tensor_{kk} = 0 = \vartheta^\tensor_{ik,k}\; $; 
and a sum over repeated indices is understood. 
The metric then takes the form
\be
  g^{ }_{\mu\nu}
  \; = \;
 a^2_{ }(\tau) \begin{pmatrix}
 -1-2h^{ }_0 & h^{ }_{\der j} \\
 h^{ }_{\der i}    & (1-2 Y)\,\delta^{ }_{ij}+
 2( \vartheta^{ }_{\der ij} + \vartheta^\tensor_{ij} ) 
 \end{pmatrix} 
 \ . \la{g_munu_scalar}
\ee
It involves four scalar degrees of freedom, 
$h^{ }_0$, $h$, $Y$, and $\vartheta$. 
Given the freedom of coordinate
transformations, only two of their linear combinations are 
physical, but we retain all four, in order to  
parametrize the ambiguity and 
to be able to reconstruct any gauge. 

For inverting 
the metric, we note that 
for a matrix of the form
\be
 g^{ }_{\mu\nu} 
 \;=\;
 a^2_{ }
 \,\bigl(\, \eta + A \,\bigr)
 \;, \quad
 \eta
 \;\equiv\;
 \biggl( 
 \begin{array}{cc}
   -\hspace*{0.3mm}1 & 0 \\ 0 & \mathbbm{1}
 \end{array} \biggr) 
 \;,
\ee
the inverse reads 
\be
 g^{\mu\nu}_{ }
 \;=\;
 \frac{1}{a^2_{ }}
 \,\bigl(\, 
    \eta
  - \eta A\hspace*{0.3mm} \eta 
  + \eta A\hspace*{0.3mm} \eta A\hspace*{0.3mm} \eta
  + \ldots
 \,\bigr)
 \;. \la{general_inverse}
\ee
Treating $h^{ }_0$, $h$, $Y$, and $\vartheta$ as first-order variables,
i.e.\ of $\rmO(\delta)$
(though they may contain second and higher-order corrections), 
we work up to second order in scalar perturbations
(i.e.\ up to and including $\rmO(A^2_{ })\sim \rmO(\delta^2_{ })$). 
The tensor perturbations are counted as second-order variables, 
i.e.\ of $\rmO(\delta^2_{ })$, 
so that we can remain at linear order in their treatment. 

Given the metric and its inverse, 
we compute the Christoffel symbols, 
$
 \Gamma^{\rho}_{\mu\nu}
 \equiv 
 {g}^{\rho\sigma}_{ } ( {g}_{\sigma\mu,\nu}^{ }
  + {g}_{\sigma\nu,\mu}^{ } - {g}_{\mu\nu,\sigma}^{ } )/2
$; 
the Ricci tensor,  
$
 R^{ }_{\mu\nu}
 \; \equiv \;
   {\Gamma}^\alpha_{\mu\nu,\alpha}
 - {\Gamma}^\alpha_{\mu\alpha,\nu}
 + {\Gamma}^\beta_{\mu\nu}  {\Gamma}^\alpha_{\beta\alpha}
 - {\Gamma}^\beta_{\mu\alpha} {\Gamma}^\alpha_{\nu\beta}
$;
the Ricci scalar,  
$
 R
 \; \equiv \;
 {g}^{\mu\nu}_{ } {R}^{ }_{\mu\nu}
$; 
and the Einstein tensor, 
$
 {G}^{ }_{\mu\nu}
 \; = \; 
 {R}^{ }_{\mu\nu}
 - {g}^{ }_{\mu\nu} {R}/2 \vphantom{\Big |}
$.
For the non-diagonal spatial part, 
the result of this computation
can be expressed as 
\ba
  G^\rmi{ }_{ij} & \overset{i\neq j}{\supset} &
 \bigl[\, 
  \partial_\tau^{\hspace*{0.2mm}2} 
 + 2 \H \partial_\tau^{ }  
 - 2 (2\H' + \H^2_{ }) 
 \,\bigr]\,
 \bigl(\,
 \vartheta^{ }_{\der ij}
 + 
 \vartheta^\rmi{t}_{ij}
 \,\bigr)
 - \nabla^2_{ } \vartheta^\rmi{t}_{ij}
 \nn[2mm]
 &  & \;-\; 
 \bigl[\,  h^{ }_0 - Y + (\partial^{ }_\tau + 2 \H ) h
 \,\bigr]^{ }_{\der ij}
 \nn[2mm]
 &  &
 \;+\;
 \G^{(h^{ }_0 \times h^{ }_0) }_{ij}
 \;+\;
 \G^{(h^{ }_0 \times Y) }_{ij}
 \;+\;
 \G^{(h^{ }_0 \times h) }_{ij}
 \;+\;
 \G^{(h^{ }_0 \times \vartheta) }_{ij}
 \nn[2mm]
 &  &
 \;+\;
 \G^{(Y \times Y) }_{ij}
 \;+\;
 \G^{(Y \times h) }_{ij}
 \;+\;
 \G^{(Y \times \vartheta) }_{ij}
 \nn[2mm]
 &  &
 \;+\;
 \G^{(h \times h) }_{ij}
 \;+\;
 \G^{(h \times \vartheta) }_{ij}
 \;+\;
 \G^{(\vartheta \times \vartheta) }_{ij}
 + 
 \ord(\delta^3_{ })
 \;, \la{Gij_t_1}
\ea
where the terms $\G^{( f \times g )}_{ij}$
are shown in \eqs\nr{G_h0_h0}--\nr{G_theta_theta}. 

Next, we introduce first-order gauge-invariant variables. 
Two independent ones exist 
(cf.\ the discussion below \eq\nr{g_munu_scalar}), 
and we choose the Bardeen potentials~\cite{Bardeen:1980kt} for this, 
defined as 
\begin{align}
 \phi &\;\equiv\; 
 h_0^{ } + (\partial^{ }_\tau + \H) (\,h - \vartheta'\,) 
 \;, 
 \la{def_phi}\\[2mm]
 \psi &\;\equiv\; 
 Y 
 + \H (\vartheta'-h)
 \;. 
 \la{def_psi}
\end{align}
In order to go over to the Bardeen potentials, 
we substitute 
$
 h^{ }_0 = \phi - (\partial^{ }_\tau + \H) (\,h - \vartheta'\,)
$
and 
$
 Y = \psi + \H (h - \vartheta')
$.
Consequently, 
the functions 
$\G^{(\phi\times \phi)}_{ij}$, 
$\G^{(\phi\times\psi)}_{ij}$, 
$\G^{(\psi\times\psi)}_{ij}$ are gauge invariant
and unambiguous 
up to second order, whereas the 
presence of either $h$ or $\vartheta$ signals ambiguity. 
We stress that the procedure represents an invertible basis transformation, 
and the original variables and equations can always be recovered 
by inserting \eqs\nr{def_phi} and \nr{def_psi}. 

The advantage of the substitutions 
below \eqs\nr{def_phi} and \nr{def_psi} is that  
$h$ and $\vartheta$ do {\em not} appear in the first-order
Einstein equations, once the right-hand side is also expressed
in terms of gauge-invariant variables. 
Therefore, $h$ and $\vartheta$ display no specific time evolution, 
and can be assigned arbitrary values at any time, whereby
they parametrize the gauge ambiguity.

For gravitational waves, 
we project out the tensor part of \eq\nr{Gij_t_1}.
By definition the tensor projector, $\mathbbm{T}^{kl}_{ij}$, 
is transverse with respect to spatial derivatives,
\be
       \mathbbm{T}^{kl}_{ij} (...)^{ }_{\der k} 
 \;=\; \mathbbm{T}^{kl}_{ij} (...)^{ }_{\der l}
 \;=\; 0
 \;. 
 \la{tmn_props}
\ee
Therefore indices can be re-ordered~\cite{rai}, 
$
 \mathbbm{T}^{kl}_{ij} \,f\, g^{ }_{\der kl}
 \; = \; 
 - 
 \mathbbm{T}^{kl}_{ij} f^{ }_{\der k}\, g^{ }_{\der l}
 \; =  \; 
 \mathbbm{T}^{kl}_{ij} f^{ }_{\der kl}\, g^{ }_{ }
$.
In the following, we denote
$
 \{ (...)^{ }_{ij} \}^\tensor_{ } 
 \; \equiv \; 
 \mathbbm{T}^{kl}_{ij} (\, ... \,)^{ }_{kl} 
$. 

Inserting \eqs\nr{def_phi} and \nr{def_psi} into \eq\nr{Gij_t_1}
and projecting with $\mathbbm{T}^{kl}_{ij}$, we find
\ba
  G^\rmi{t}_{ij} 
 & 
  = 
 &
 \; - \, 
 \bigl[\, 
 a^2_{ } \hspace*{0.3mm}
 \raise-0.2ex\hbox{$\square$} \hspace*{0.3mm}
 \,+\,
 2 (2\H' + \H^2_{ })
 \,\bigr]\,
 \vartheta^\tensor_{ij} 
 \; + \; 
 \bigl\{\,
 \phi \hspace*{0.3mm} \phi^{ }_{\der ij}
 + 
 (2 \phi - \psi) \psi^{ }_{\der ij}
 \,\bigr\}^\tensor_{ }
 \nn[3mm]
 &  & \;+\,\biggl\{\,
 (\phi' + 3 \psi' + 4\H \phi) h^{ }_{\der ij}
 \nn[3mm]
 &  & \hspace*{4mm} \;+\, 
 4 [\H \phi' + ( 2 \H' + \H^2_{ } )\phi]
 \hspace*{0.3mm} \vartheta^{ }_{\der ij}
 - 
 (\phi' + 4 \H \phi)\hspace*{0.3mm} 
    \vartheta^{\hspace*{0.2mm}\prime }_{\der ij} 
 + \phi^{ }_{\der k} \hspace*{0.3mm} \vartheta^{ }_{\der ijk}
 \nn[3mm]
 &  & \hspace*{4mm} \;+\, 
 4 (\psi'' + 2 \H \psi') \hspace*{0.3mm}  \vartheta^{ }_{\der ij}
 - 3 \psi' \hspace*{0.3mm}  \vartheta^{\hspace*{0.2mm}\prime }_{\der ij}
 -   \hspace*{0.3mm}
     \psi^{ }_{\der k} \hspace*{0.3mm} \vartheta^{ }_{\der ijk}
 + 2 \hspace*{0.3mm}
     ( \phi^{ }_{\der kk} 
 -  
     \psi^{ }_{\der kk} )
   \hspace*{0.3mm} \vartheta^{ }_{\der ij}
 \nn[3mm]
 &  & \hspace*{4mm} \;+\, 
 2(\H' - \H^2_{ })
   ( h 
   - \vartheta^{\hspace*{0.2mm}\prime }_{ }
    ) 
    ( 
     h
    - \vartheta^{\hspace*{0.2mm}\prime }_{ }
    )^{  }_{\der ij}
 + 4 (\H'' - \H \H' - \H^3_{ })
   ( h - \vartheta^{\hspace*{0.2mm}\prime }_{ } )\hspace*{0.3mm}
         \vartheta^{ }_{\der ij}
 \nn[3mm]
 &  & \hspace*{4mm} \;+\, 
 \frac{a^2_{ }  \raise-0.2ex\hbox{$\square$} }{2}
 \bigl[\,
   ( h 
   - \vartheta^{\hspace*{0.2mm}\prime }_{ }
   - 4 \H \vartheta
    ) 
    ( 
     h
    - \vartheta^{\hspace*{0.2mm}\prime }_{ }
    )^{  }_{\der ij}
 - 4 \psi \hspace*{0.3mm} \vartheta^{ }_{\der ij} 
 - \vartheta^{ }_{\der k} \hspace*{0.3mm} 
   \vartheta^{ }_{\der ijk}
 \,\bigr]
 \,\biggr\}^\rmi{t}_{ }
 \;+\, 
 \ord(\delta^3_{ })
 \;. \hspace*{7mm} \la{Gij_t_2}
\ea
Here the d'Alembert operator acts in conformal coordinates as 
\be
 a^2_{ } \hspace*{0.3mm}
 \raise-0.2ex\hbox{$\square$} \hspace*{0.3mm} Q \; \equiv \; 
 \bigl(\, - \partial^2_\tau - 2 \H \partial^{ }_\tau
  + \nabla^2_{ } \,\bigr)\hspace*{0.3mm} Q 
 \;, \la{dalembert}
\ee
and can in practical computations be identified as the combination
\be
 a^2_{ } \hspace*{0.3mm}
 \raise-0.2ex\hbox{$\square$} \hspace*{0.3mm} (f \hspace*{0.3mm} g)
 \; = \; 
 - f'' \hspace*{0.3mm} g 
 - 2 f' \hspace*{0.3mm} g'
 - f \hspace*{0.3mm} g'' 
 - 2 \H \hspace*{0.3mm} ( f' \hspace*{0.3mm} g + f \hspace*{0.3mm} g')
 + f^{ }_{\der kk} \hspace*{0.3mm} g^{ }_{ }
 + 2 f^{ }_{\der k} \hspace*{0.3mm} g^{ }_{\der k}
 + f^{ }_{ } \hspace*{0.3mm} g^{ }_{\der kk}
 \;. 
\ee

%
\subsection{Right-hand side of the Einstein equations}
\la{ss:rhs}

Moving to the right-hand side of the Einstein equations, 
we expand $T^{ }_{\mu\nu}$ from \eq\nr{Tmunu_mixed}
up to second order in scalar perturbations. 
For this we write the scalar quantities as 
\be
  \varphi 
  \; \overset{\rmii{ }}{\equiv} \; 
  \bar\varphi + \delta\varphi
  \;, \quad
  p^{ }_n 
  \; \overset{\rmii{ }}{\equiv} \; 
  \bar p^{ }_n + \delta p^{ }_n
  \;, \quad
  e^{ }_n
  \; \overset{\rmii{ }}{\equiv} \; 
  \bar e^{ }_n + \delta e^{ }_n
  \;,
  \la{2nd_ord_perts}
\ee
which can be viewed as definitions of 
$
 \delta\varphi
$, 
$ 
 \delta p^{ }_n
$
and 
$
 \delta e^{ }_n
$.
The background values ($\bar\varphi$, $\bar p^{ }_n$, $\bar e^{ }_n$) 
are set to 
be independent of the spatial coordinates. 
We denote 
$\bar p \equiv \sum_n \bar p^{ }_n$, 
$\delta p \equiv \sum_n \delta p^{ }_n$, 
$\bar e \equiv \sum_n \bar e^{ }_n$, and 
$\delta e \equiv \sum_n \delta e^{ }_n$.  
The scalar (or irrotational)
velocity perturbation, $\delta v^{ }_n$, needs to be defined so that
the normalization condition 
$u^{\vphantom{\mu} }_{n \mu} u^\mu_{n\vphantom{\mu}} = - 1$ 
is satisfied $\forall n$, 
which leads to  
\be
  u_{n \mu}^{\vphantom{\mu}}   
  \; \underset{\rmii{ }}{\overset{\rmii{ }}{=}} \;
  a(-1-h_0,h^{ }_{,i} - \delta v^{ }_{n,i}) + \ord(\delta^2_{ })  
 \;, \quad
 n \;\ge\; 1
 \;. 
 \la{def_delta_v}
\ee
The anisotropic (shear) stress, $\Pi^{ }_{\mu\nu}$, 
is assumed spatial and traceless, but is otherwise kept as such, 
though at the end we find it natural to consider 
it as a {\em second-order} quantity, $\sim\rmO(\delta^2_{ })$,
just like $\vartheta^\tensor_{ij}$. 
It is decomposed into scalar and tensor parts
like $\vartheta^{ }_{ij}$ in \eq\nr{scalar}, 
\be
 \Pi^{ }_{ij}
 \; \equiv \;
 \Pi^\tensor_{ij}
 + 
 \Pi^{ }_{\der ij}
 - \frac{\delta^{ }_{ij}}{3}\, \Pi^{ }_{\der kk} 
 \;. \la{Pi}
\ee

Inserting \eqs\nr{2nd_ord_perts} and \nr{def_delta_v}
as well as the metric from \eq\nr{g_munu_scalar}
into \eq\nr{Tmunu_mixed}, and expanding to second order, 
the components
contributing to the $i\neq j$ part of $T^{ }_{ij}$ become 
\ba
 \varphi^{ }_{,i}\, \varphi^{ }_{,j}
 & 
 \overset{\rmii{\nr{2nd_ord_perts}}}{=} 
 & 
 \delta\varphi^{ }_{,i}\, \delta\varphi^{ }_{,j}
 \;, \la{phi_prime_t} \\[3mm]
 - \frac{ g^{ }_{ij}\, 
 \varphi^{ }_{,\alpha} \varphi^{,\alpha}_{ } }{2} 
 & \underset{i\;\neq\;j}
  {\overset{\rmii{\nr{g_munu_scalar}}}{=}} & 
  (\vartheta^\tensor_{ij} + \vartheta^{ }_{\der ij}) 
  \biggl[\, 
  (1 - 2 h^{ }_0) (\bar{\varphi}')^2 
  + 2 {\bar\varphi}'\delta\varphi'
  \,\biggr]
 + \ord(\delta^3_{ })
 \;, \la{phi_t_diag} \\[3mm]
 ( e^{ }_n + p^{ }_n ) \,  u^{ }_{n i}u^{ }_{n j}
 & 
  \underset{n \, \ge \, 1}{
  \overset{\rmii{\nr{2nd_ord_perts}}}{=}} 
 &
  a^2_{ }(\bar{e}^{ }_n + \bar{p}^{ }_n)
 (h-\delta v^{ }_n)^{ }_{,i}
 (h-\delta v^{ }_n)^{ }_{,j}
 + \ord(\delta^3_{ })
 \;, \la{uu_t} \\[3mm]
 p^{ }_n \, g^{ }_{ij}
 & \underset{i\;\neq\;j}
  {\overset{\rmii{\nr{g_munu_scalar}}}{=}} & 
 2 a^2_{ }
   (\vartheta^\tensor_{ij} + \vartheta^{ }_{\der ij}) 
 \bigl( \bar{p}^{ }_n + \delta p^{ }_n \bigr)
 \;. \la{p_t_diag}
\ea

The scalar perturbations introduced in \eq\nr{2nd_ord_perts}
are not gauge invariant. We employ  
curvature perturbations for capturing gauge-invariant
matter fluctuations. They are defined as  
\ba
 \R^{ }_\varphi
 &
  \underset{\rmii{ }}
 {\overset{\rmii{ }}{\equiv}} 
 & 
 -\, Y 
  - \H\,\frac{\delta\varphi}{\bar\varphi\hspace*{0.3mm}{}'}
 \la{R_varphi}
 \;, \\[2mm] 
 \R^{ }_{n v}
 &
  \underset{\rmii{ }}
 {\overset{\rmii{ }}{\equiv}} 
 & 
 -\, Y 
 + \H \, (h - \delta v^{ }_n)
 \;, \quad n \; \ge \; 1
 \;, \la{R_v}  
\ea
where $Y = h^{ }_\rmii{D} + \vartheta^{ }_{\der kk}/3$ 
is the shorthand from \eq\nr{scalar}.  
One can similarly define a curvature perturbation for
the energy density, 
$
 \R^{ }_{n e}
 \equiv  
  -\, Y 
  - \H\, {\delta e^{ }_n} / {\bar e_n^{\hspace*{0.3mm}{}\prime} }
$,
and this is what is frequently denoted by $\zeta$
in the literature.  
When the first-order
Einstein equations are written in terms of
$\R^{ }_\varphi$, $\R^{ }_{n v}$, $\R^{ }_{n e}$, 
and the Bardeen potentials $\phi$ and $\psi$, we obtain
a closed set of stochastic differential equations, 
in which the remaining metric perturbations
$h$ and $\vartheta$ do not appear~\cite{dissip}. 
The stochastic-noise autocorrelators are determined by
the local temperature~\cite{fluctu}.

With the help of \eqs\nr{R_varphi} and \nr{R_v}, 
we can substitute 
\ba
 \delta\varphi^{ }_{\der i}
 \,\delta\varphi^{ }_{\der j}
 & 
  \overset{\rmii{\nr{phi_prime_t}}}{
  \underset{\rmii{\nr{R_varphi}}}{=}}  
 & 
 \frac{(\bar\varphi\hspace*{0.3mm}')^2_{ }}{\H^2}
 (\R^{ }_\varphi + Y )^{ }_{\der i}
 (\R^{ }_\varphi + Y )^{ }_{\der j}
 \;, \la{Tij_varphi} 
 \\[2mm] 
 ( h - \delta v^{ }_n)^{ }_{\der i}
 ( h - \delta v^{ }_n)^{ }_{\der j}
 & 
  \overset{\rmii{\nr{uu_t}}}{
  \underset{\rmii{\nr{R_v}}}{=}}  
 & 
 \frac{1}{\H^2}
 (\R^{ }_{n v} + Y )^{ }_{\der i}
 (\R^{ }_{n v} + Y )^{ }_{\der j}
 \;. \la{Tij_v} 
\ea
In order to simplify the outcome, we can make use of lower-order
Einstein equations. The background (Friedmann) equations read
\ba
 8\pi G \bigl[\, (\bar{\varphi}\hspace*{0.3mm}')^2_{ } 
 + 2 a^2_{ }\bar{p} \,\bigr]
 &
 \overset{\rmii{ }}{=}
 & 
 -2(2\H' + \H^2_{ })
 \;, \la{bg_1}
 \\[2mm]
 8\pi G \bigl[\, 
 (\bar{\varphi}\hspace*{0.3mm}')^2_{ }
 +
 a^2_{ }(\bar{e} + \bar{p})
 \,\bigr]
 &
 \overset{\rmii{ }}{=}
 & 
 2(\H^2_{ } - \H') 
 \;. \la{bg_2}
\ea
In addition, we can make use of first-order relations. 
The components $0i$, $i\neq j$, and $i=j$ of 
the Einstein equations yield, respectively, 
\ba
  \H h^{ }_0 
 +   
   Y'
 & = & 
 4\pi G \bigl[\, 
  a^2_{ }
 {\textstyle \sum_n } 
 ( \bar{e}^{ }_n + \bar{p}^{ }_n) (\delta v^{ }_n - h)
  + {\bar\varphi}' \delta\varphi
  \,\bigr]
 \;, \la{einstein_0i}
 \\[3mm]
   h^{ }_0 
 + (\partial^{ }_\tau + 2 \H) (h - \vartheta')
 - Y 
 & = &
 -\,
  8\pi G a^2_{ }\barpPi 
 \;, \la{einstein_ij_d}
 \\[3mm]
 & & \hspace*{-6.5cm} 
 \bigl[\, 
   2 (2\H' + \H^2_{ }) 
  + 2 \H \partial^{ }_\tau 
  + \nabla^2_{ } \,\bigr] h^{ }_0
 \; + \; 
 \bigl[\,
 2 (\partial^2_\tau + 2 \H \partial^{ }_\tau) - \nabla^2_{ }
 \,\bigr]
 \hspace*{0.3mm} Y
 \; + \;
 (\partial^{ }_\tau + 2 \H) \nabla^2_{ } (h - \vartheta') 
 \nn[3mm]
 & = &  
 8\pi G  
 \biggl[ a^2_{ } \biggl( \delta p 
   - \frac{\nabla^2_{ }\barpPi}{3} \biggr)
  +  {\bar\varphi}' \bigl( \delta\varphi'
  - h^{ }_0 \bar\varphi\hspace*{0.3mm}{}' \bigr)  
 \biggr]
 \;. \label{einstein_ij_t}
\ea
Eq.~\nr{einstein_0i} can be used for expressing the combination
$ 
  (\bar{\varphi}\hspace*{0.3mm}')^2_{ } \R^{ }_\varphi 
  + a^2_{ }\sum_n (\bar e^{ }_n + \bar p^{ }_n) \R^{ }_{n v}
$
in terms of metric perturbations, 
and 
\eq\nr{einstein_ij_t}
allows us to eliminate $\delta p = \sum_n \delta p^{ }_n$.

After having eliminated $\delta\varphi$, 
$\delta v^{ }_n$ and $\delta p^{ }$ 
in favour of $\R^{ }_\varphi$, $\R^{ }_{n v}$, and metric perturbations, 
we replace $h^{ }_0$ and $Y$ by Bardeen potentials, 
as specified below \eq\nr{def_psi}.  
Subsequently, we project $T^{ }_{ij}$ to the tensor channel.
Putting everything together, this yields
\ba
 8\pi G\, T^\rmi{t}_{ij} 
 & 
 = 
 & 
 -  2 (2\H' + \H^2_{ })
 \,\vartheta^\rmi{t}_{ij}
 + 
 8 \pi G \,  a^2_{ }\Pi^\rmi{t}_{ij}
 \;+\,
  \biggl\{\,
 2\hspace*{0.3mm}
 \biggl[\, 2 \phi 
 + \biggl( 1 - \frac{\H'}{\H^2_{ }} \biggr) \psi
 + \frac{2}{\H} \psi'
 \,\biggl] \hspace*{0.3mm} \psi^{ }_{\der ij}
 \,\biggr\}^\tensor_{ }
 \nn[3mm]
 &  & \;+\, 
 \frac{8\pi G}{\H^2_{ }}
 \bigl\{\, 
   (\bar{\varphi}\hspace*{0.3mm}')^2_{ } 
   \R^{ }_{\varphi\der i}
   \R^{ }_{\varphi\der j}
  + 
   a^2_{ }\,{\textstyle \sum_n} (\bar e^{ }_n + \bar p^{ }_n)
   \R^{ }_{n v\der i}
   \R^{ }_{n v\der j}
  \,\bigr\}^\tensor_{ }
 \nn[3mm]
 &  & 
 \;+\, 
 \biggl\{\;
  4 \hspace*{0.3mm} 
   (
     \psi'  + \H \phi
   )
   \hspace*{0.3mm} 
   ( h - \vartheta^{\hspace*{0.2mm}\prime }_{ } ) ^{ }_{\der ij}
 \;+\,
 4 [\H \phi' + ( 2 \H' + \H^2_{ } )\phi]
  \hspace*{0.3mm} \vartheta^{ }_{\der ij}
 \nn[3mm]
 &  & \hspace*{4mm} \;+\, 
 4 (\psi'' + 2 \H \psi') \hspace*{0.3mm}  \vartheta^{ }_{\der ij}
 + 2 \hspace*{0.3mm}
   \biggl( \phi
  - 
     \psi
  + \frac{8\pi G a^2_{ }\Pi }{3}
   \biggr)^{ }_{\der kk} 
   \hspace*{0.3mm} \vartheta^{ }_{\der ij}
 \nn[3mm]
 &  & \hspace*{4mm} \;+\, 
 2(\H' - \H^2_{ })
   ( h 
   - \vartheta^{\hspace*{0.2mm}\prime }_{ }
    ) 
    ( 
     h
    - \vartheta^{\hspace*{0.2mm}\prime }_{ }
    )^{  }_{\der ij}
 + 4 (\H'' - \H \H' - \H^3_{ })
   ( h - \vartheta^{\hspace*{0.2mm}\prime }_{ } )\hspace*{0.3mm}
         \vartheta^{ }_{\der ij} 
 \;\biggr\}^\rmi{t}_{ }
 \nn[3mm]
 & &  \hspace*{4mm}
 \;+\; 
 \ord(\delta^3_{ })
 \;.  \la{Tij_t_2}
\ea
The first two rows of \eq\nr{Tij_t_2} are manifestly gauge invariant, 
whereas the remaining rows contain the unconstrained fields 
$h$ and $\vartheta$, and are therefore ambiguous. 

%
\subsection{Equating the sides}
\la{ss:equate}

The remaining step is to equate \nr{Gij_t_2} and \nr{Tij_t_2}.
Moving most of the terms
to the right-hand side, we observe an almost complete cancellation 
of the gauge-dependent parts of \eq\nr{Tij_t_2}. 
Left over is an effective wave equation, reading  
\ba
 -\;
 \raise-0.2ex\hbox{$\square$} \hspace*{0.3mm}
 \vartheta^\rmi{t}_{ij}
 & 
  \overset{\rmii{\nr{Gij_t_2}}}
  {\underset{\rmii{\nr{Tij_t_2}}}{=}} 
 & 
 8 \pi G \, \Pi^\rmi{t}_{ij}
 \;+\, 
 \frac{6}{a^2_{ }}
 \biggl\{\, 
 \frac{ 
   (\bar{\varphi}\hspace*{0.3mm}')^2_{ } 
   \R^{ }_{\varphi\der i}
   \R^{ }_{\varphi\der j}
  + 
   a^2_{ }\sum_n (\bar e^{ }_n + \bar p^{ }_n )
   \R^{ }_{n v\der i}
   \R^{ }_{n v\der j}
 }{
  (\bar{\varphi}\hspace*{0.3mm}')^2_{ } 
 + 2 a^2_{ }\bar{e}
 }
 \,\biggr\}^\tensor_{ }
 \nn[3mm]
 &  & \;+\,
 \frac{1}{a^2_{ }}
 \biggl\{\;
   \phi \hspace*{0.3mm} ( 2 \psi - \phi)^{ }_{\der ij} 
 + \biggl( 3 - \frac{2 \H'}{\H^2_{ }} \biggr)
   \psi \hspace*{0.3mm} \psi^{ }_{\der ij}
 + \frac{4}{\H} \hspace*{0.3mm} \psi^{\hspace*{0.2mm}\prime }_{ }
                \hspace*{0.3mm} \psi^{ }_{\der ij}
 \;\biggr\}^\tensor_{ }
 \nn[3mm]
 &  & 
  \;+\, 
 \frac{1}{a^2_{ }}
 \biggl\{\;
   (
     \psi  - \phi
   )^{\prime}_{ }
   \hspace*{0.3mm} 
   ( h - \vartheta^{\hspace*{0.2mm}\prime }_{ }) ^{ }_{\der ij}
 + ( \psi - \phi)^{ }_{\der k} \hspace*{0.3mm} \vartheta^{ }_{\der ijk}
 + \frac{16 \pi G a^2_{ }\Pi_{\der kk} 
   \hspace*{0.3mm} \vartheta_{\der ij}
   }{3}
 \;\biggr\}^\rmi{t}_{ }
 \nn[3mm]
 & &  
 - \; \frac{1}{2} \hspace*{0.4mm}
 \raise-0.2ex\hbox{$\square$} \hspace*{0.3mm}
 \bigl\{\,
   ( h 
   - \vartheta^{\hspace*{0.2mm}\prime }_{ }
   - 4 \H \vartheta
    ) 
    ( 
     h
    - \vartheta^{\hspace*{0.2mm}\prime }_{ }
    )^{  }_{\der ij}
 - 4 \psi \hspace*{0.3mm} \vartheta^{ }_{\der ij} 
 - \vartheta^{ }_{\der k} \hspace*{0.3mm} 
   \vartheta^{ }_{\der ijk}
 \,\bigr\}^\tensor_{ }
 \;+\; 
 \ord(\delta^3_{ })
 \;. \hspace*{6mm}  \la{sigw}
\ea
Here we made use of 
$
 8\pi G/\H^2_{ } = 6 / 
 [\,
  (\bar{\varphi}\hspace*{0.3mm}')^2_{ } 
 + 2 a^2_{ }\bar{e}
 \,]
$, 
obtained from \eqs\nr{bg_1} and \nr{bg_2}.
The first two rows on the right-hand side 
of \eq\nr{sigw} are gauge invariant, 
the last two not manifestly so, because they contain the 
unconstrained functions $h$ and $\vartheta$ 
(however see the discussion leading to \eq\nr{sigw_2}). 

We note that on the last row of \eq\nr{sigw},  
$
 -4 [
    \H (      h
    - \vartheta^{\hspace*{0.2mm}\prime }_{ } )
    + 
    \psi 
    ]
    \, \vartheta^{ }_{\der ij}
 = 
 - 4 \hspace*{0.3mm} Y \hspace*{0.3mm} \vartheta^{ }_{\der ij}
$
according to \eqs\nr{scalar} and \nr{def_psi}.
Comparing with literature, 
our variable $-2Y$ corresponds to what is denoted by $\psi$ 
in ref.~\cite{counterterm2}; 
our $ h 
   - \vartheta^{\hspace*{0.2mm}\prime}_{ } $
corresponds to their $\sigma$; 
and our $2 \vartheta$ to their $E$.
Then the last row agrees with 
\eq(27) of ref.~\cite{counterterm2}
(even though their $T^\tensor_{ij}$ 
did not contain $\varphi$).  
This provides for a non-trivial crosscheck of \eq\nr{sigw}.

%
\section{Physical interpretation of the effective wave equation}
\la{se:interp}

Let us elaborate on a few key properties of \eq\nr{sigw}, and on how
it can be turned into an unambiguous evolution equation.  

%
\paragraph{Gauge dependence.}

If $\phi \neq \psi$ or $\Pi \neq 0$, the third row of \eq\nr{sigw}  
depends on the functions $h$ and $\vartheta$. 
The dynamics of $h$ and $\vartheta$ are not fixed by the 
first-order Einstein equations, once we have expressed the latter
in terms of the Bardeen potentials $\phi$, $\psi$ and the 
curvature perturbations $\R^{ }_\varphi$, $\R^{ }_{nv}$ and $\R^{ }_{ne}$. 
This leads to an ambiguity. 
The ambiguity drops
out if $\phi - \psi \sim \rmO(\delta^2_{ })$
and $\Pi \sim \rmO(\delta^2_{ })$. Expressing 
\eq\nr{einstein_ij_d} in terms of the Bardeen potentials
from \eqs\nr{def_phi} and \nr{def_psi}, 
it reads 
$ 
 \psi -\phi 
 =
 8\pi G a^2_{ }\barpPi
$.
We thus see that a sufficient condition for 
the ambiguity to drop out is that $\Pi \sim \rmO(\delta^2_{ })$,  
i.e.\ that the anisotropic stress is small enough to be counted
as a {\em second-order quantity}. Then it does not 
appear in the first-order Einstein equations, but its 
tensor part still plays a role (see below). 

The second apparent source of ambiguity in 
\eq\nr{sigw} originates from the last row. 
We note that this term affects the 
gravitational wave amplitude through its value
at the observation time, rather than in the early universe. 
Concretely, denoting
by $G^{ }_\rmii{R}(\X;\Y) = \square^{-1}_{ }$ a retarded Green's function, 
the wave equation 
\be
 -\, \raise-0.2ex\hbox{$\square$} \hspace*{0.3mm} \vartheta^\tensor_{ij}(\X) 
 \; = \;
 \mathcal{S}^\tensor_{ij}(\X)
 -\, \raise-0.2ex\hbox{$\square$} \hspace*{0.3mm} \Q^\tensor_{ij}(\X)
\ee
has the solution
\be
 \vartheta^\tensor_{ij}(\X)
 - 
 \Q^\tensor_{ij}(\X) 
 \; = \; 
 - 
 \int_\Y  G^{ }_\rmii{R}(\X;\Y)\, \mathcal{S}^\tensor_{ij}(\Y)
 \;. \la{soln}
\ee
The combination $ 2(\vartheta^\rmi{t}_{ij} - \Q^\tensor_{ij}) $
corresponds to what is denoted by $H^N_{ij}$ in \eqs(27) and~(69)
of ref.~\cite{counterterm2}. It is shown to be gauge invariant
at second order in their appendix~B 
(building upon ref.~\cite{gauge2}). Therefore, \eq\nr{soln}
represents a gauge-invariant solution.

However, though restoring gauge invariance, 
$ \Q^\tensor_{ij} $ is still ambiguous, because it depends on 
the unconstrained functions $h$ and $\vartheta$. 
As $h$ and $\vartheta$ are non-propagating, we are free
to set their values to zero at the observation time, 
in order to carry out the measurement 
in a locally flat region, 
defined in the sense of the equivalence principle. 
This appears to  
promote $\vartheta^\tensor_{ij}$ to a physical quantity
at the second order. 
That said, the definition of a precise observable
goes beyond the scope of the present study
(for recent work, see ref.~\cite{Domenech:2025ccu}).

We remark in passing that in some of the literature, 
when non-Newtonian gauges are considered, 
the functions appearing in $\Q^\tensor_{ij}$ satisfy evolution 
equations, and show a definite time dependence
(cf.,\ e.g.,\ ref.~\cite{counterterm1}). 
The reason is that the first-order evolution equations had not
been expressed in terms of the Bardeen potentials and curvature
perturbations, as we do here. In that situation, physical 
information is contained in the difference
$
 \vartheta^\tensor_{ij}
 - 
 \Q^\tensor_{ij} 
$.

To summarize, the physical content of \eq\nr{sigw} can be 
captured by solving for the difference
$ \vartheta^\rmi{t}_{ij} - \Q^\tensor_{ij} $, from 
\ba
 -\;
 \raise-0.2ex\hbox{$\square$} \hspace*{0.3mm}
 (\, \vartheta^\rmi{t}_{ij} - \Q^\tensor_{ij} \,)
 \; 
  \overset{\rmii{\nr{sigw}}}
  {\underset{\phi\;\approx\;\psi}{ = }} 
 \;
 8 \pi G \, \Pi^\rmi{t}_{ij}
 &
 \,+\, 
 & 
 \frac{6}{a^2_{ }}
 \biggl\{\, 
 \frac{ 
   (\bar{\varphi}\hspace*{0.3mm}')^2_{ } 
   \R^{ }_{\varphi\der i}
   \R^{ }_{\varphi\der j}
  + 
   a^2_{ }
  \sum_n
  (\bar e^{ }_n + \bar p^{ }_n)
   \R^{ }_{n v\der i}
   \R^{ }_{n v\der j}
 }{
  (\bar{\varphi}\hspace*{0.3mm}')^2_{ } 
 + 2 a^2_{ }\bar{e}
 }
 \,\biggr\}^\tensor_{ }
 \hspace*{1mm}
 \nn[3mm]
 &
 \,+\,
 & 
 \frac{2}{a^2_{ }}
 \biggl\{\;
   \psi^{ }_{\der ij} 
    \hspace*{0.3mm} 
   \biggl( \frac{2 \partial^{ }_\tau }{\H}
  +  2 - \frac{\H'}{\H^2_{ }} \biggr)
   \psi
 \;\biggr\}^\tensor_{ }
 \;, 
 \la{sigw_2}
\ea
where the approximation refers to $\phi = \psi + \rmO(\delta^2_{ })$.
We remark that circumstances can be imagined
where the anisotropic stress and thus $\phi-\psi$ 
are not small, like when the hydrodynamic expansion is about to break 
down due to large gradients. However, in such a situation the hydrodynamic
expansion should be replaced by a more microscopic treatment in any case, 
so this does not represent a substantial restriction on the 
generality of \eq\nr{sigw_2}.  

%
\paragraph{Structure of the effective source term.}

The first row of \eq\nr{sigw_2} contains
the curvature perturbations associated with a scalar field 
($\R^{ }_\varphi$, cf.\ \eq\nr{R_varphi}) and fluid velocities
($\R^{ }_{n v}$, cf.\ \eq\nr{R_v}). The terms are multiplied by 
the respective background enthalpy densities, 
$ 
 (\bar{\varphi}\hspace*{0.3mm}')^2_{ } 
$
and
$
 a^2_{ }(\bar e^{ }_n + \bar p^{ }_n)
$, 
and divided by (twice) the overall energy density. 
The (stochastic) 
evolution equations for the curvature perturbations
can be found in \eqs(3.42--44) of ref.~\cite{dissip}
(for one fluid), while the 
evolution equation for the Bardeen potential $\psi$ follows from 
\eq\nr{einstein_0i}, 
\be
 \biggl(\,
  \frac{\partial^{ }_\tau }{\H}
  +  2 - \frac{\H'}{\H^2_{ }} 
 \,\biggr)
 \, \psi 
 \;
 \underset{\phi\;\approx\;\psi}{
 \overset{\rmii{\nr{einstein_0i}}}{=}}
 \; 
 -\,3 \; 
 \frac{ 
   (\bar{\varphi}\hspace*{0.3mm}')^2_{ } 
   \R^{ }_{\varphi}
  + 
   a^2_{ }
   \sum_n
   (\bar e^{ }_n + \bar p^{ }_n)
   \R^{ }_{n v}
 }{
  (\bar{\varphi}\hspace*{0.3mm}')^2_{ } 
 + 2 a^2_{ }\bar{e}
 }
 \;. \la{evo_psi} 
\ee

Under certain conditions, \eqs\nr{sigw_2} and \nr{evo_psi} can 
be simplified. If we consider modes outside of the Hubble horizon, 
then all curvature perturbations take the same value, 
i.e.\ $\R^{ }_\varphi \approx \R^{ }_{n v}$ $\forall n$. In this case, we can 
use \eq\nr{evo_psi} to express their common value in terms of $\psi$.
Substituting into \eq\nr{sigw_2}, the right-hand side is a quadratic
function of $\psi$ and its derivatives (cf.\ \eq\nr{lit_1}). 
However, as the modes 
re-enter inside the Hubble horizon, $\R^{ }_\varphi$ and $\R^{ }_{n v}$
start to undergo acoustic oscillations, 
and they rapidly differ from each other~\cite{dissip}, 
making this simplification not viable in general. 


%
\paragraph{Comparison with literature.}

\rev{
In most of the literature, a system with a single  
curvature perturbation has been considered
(cf.,\ e.g.,\ ref.~\cite{sigw}). 
More recently, this has been generalized to 
a setup with two components, 
a radiation and a matter fluid~\cite{chicago}. Here we illustrate
how the corresponding results can be recovered from 
\eqs\nr{sigw_2} and \nr{evo_psi}. For this purpose we 
rename the right-hand side of \eq\nr{sigw_2} as 
\be
  -\;
 \raise-0.2ex\hbox{$\square$} \hspace*{0.3mm}
 (\, \vartheta^\rmi{t}_{ij} - \Q^\tensor_{ij} \,)
 \; 
  \overset{\rmii{\nr{sigw_2}}}
  {\underset{ }{\equiv}} 
 \;
 8 \pi G \, \Pi^\rmi{t}_{ij}
 \,+\, 
 \frac{S^{ }_{ij}}{a^2_{ }}
 \;, \la{sigw_3}
\ee
noting that literature
definitions of $S^{ }_{ij}$ 
differ with respect to the overall
sign and a factor 4. 

To recover the single curvature perturbation result, 
let us start by setting 
$
 (\bar{\varphi}\hspace*{0.3mm}')^2_{ } \to 0 
$
and restricting to one fluid
($n^{ }_{\rm tot} = 1$).
This choice is made for illustrative purposes, and corresponds
to a situation late after inflation.  
Furthermore, we define an equation-of-state parameter
as 
$
 w \equiv \bar p / \bar e
$.
From the background identities in \eqs\nr{bg_1} and \nr{bg_2}, 
it follows that 
$
 1 - \H'/\H^2_{ } = 3(1+w)/2
$
and
$
 - \H'/\H^2_{ } = (1+3 w)/2
$.
Equation~\nr{sigw_2} implies that 
\be
 S^{ }_{ij} 
 \;
 \overset{(\bar{\varphi}\hspace*{0.3mm}')^2_{ } \,\to\, 0}{
 \underset{n^{ }_{\rm tot} \,=\, 1}{=}}
 \;
 \biggl\{
 3 (1 + w) \R^{ }_{v,i}\R^{ }_{v,j}
 -2 \psi^{ }_{,i} 
 \biggl( \frac{\psi^{\hspace*{0.3mm}\prime}_{,j}}{\H} + \psi^{ }_{,j} \biggr)
 -2 \psi^{ }_{,j} 
 \biggl( \frac{\psi^{\hspace*{0.3mm}\prime}_{,i}}{\H} + \psi^{ }_{,i} \biggr)
 + \frac{2\H' \psi^{ }_{,i} \psi^{ }_{,j} }{\H^2_{ }}
 \biggr\}^\tensor_{ }
 \;, \hspace*{5mm} \la{lit_1_pre} 
\ee
} \rev{
\noindent
whereas \eq\nr{evo_psi} can be rewritten as 
\be
 \frac{\psi^{\hspace*{0.3mm}\prime}_{ }}{\H} + \psi 
 \;
 \overset{(\bar{\varphi}\hspace*{0.3mm}')^2_{ } \,\to\, 0}{
 \underset{n^{ }_{\rm tot} \,=\, 1}{=}}
 \;
 -\,\frac{3}{2} (1+w) \bigl(\, \R^{ }_v \, + \, \psi \,\bigr)
 \;. \la{psi_1_pre}
\ee

We now write the product appearing in the first term of 
\eq\nr{lit_1_pre} as
\be
 \R^{ }_{v,i} \R^{ }_{v,j}
 \; = \; 
 \bigl(\, \R^{ }_v \, + \, \psi \bigr)^{ }_{,i} 
 \bigl(\, \R^{ }_v \, + \, \psi \bigr)^{ }_{,j}
 \; - \; 
 \psi^{ }_{,i} 
 \bigl(\, \R^{ }_v \, + \, \psi \bigr)^{ }_{,j}
 \; - \; 
 \psi^{ }_{,j} 
 \bigl(\, \R^{ }_v \, + \, \psi \bigr)^{ }_{,i}
 \; + \; 
 \psi^{ }_{,i}\, \psi^{ }_{,j}
 \;. 
\ee
If we consider times late after inflation, 
we can safely assume $w > -1$, 
and insert then \eq\nr{psi_1_pre} for 
$\R^{ }_v + \psi$. The mixed terms cancel, leaving over
\be
 S^{ }_{ij}
 \; 
 \overset{(\bar{\varphi}\hspace*{0.3mm}')^2_{ } \,\to\, 0}{
 \underset{n^{ }_{\rm tot} \,=\, 1}{=}}
 \; 
 \biggl\{ 
 \frac{4}{3(1+w)}
 \biggl( \frac{\psi^{\hspace*{0.3mm}\prime}_{,i}}{\H} + \psi^{ }_{,i} \biggr)
 \biggl( \frac{\psi^{\hspace*{0.3mm}\prime}_{,j}}{\H} + \psi^{ }_{,j} \biggr)
 \; + \; 
 \underbrace{
 \biggl[
  3(1+w) \, + \, \frac{2\H'}{\H^2_{ }}
 \biggr]
 }_{3 + \cancel{3 w} - 1 - \cancel{3w} \; = \; 2 }
 \, 
 \psi^{ }_{,i}\, \psi^{ }_{,j} 
 \biggr\}^\tensor_{ }
 \;. \la{lit_1}
\ee
This agrees with \eq(12) of ref.~\cite{sigw}
in the limit $\psi\approx\phi$, 
and with very many other references.

The problem becomes richer if there are two different
fluids ($n^{ }_{\rm tot} = 2$), because this implies 
the presence of isocurvature modes. Being pedantic with the notation, 
the background identities from \eqs\nr{bg_1} and \nr{bg_2} read 
$
 1 - \H'/\H^2_{ } = 
 3( 
  \bar e^{ }_1 + \bar e^{ }_2 + \bar p^{ }_1 + \bar p^{ }_2
 )/[2( \bar e^{ }_1 + \bar e^{ }_2)]
$
and
$
 - 2 \H'/\H^2_{ } = 
 [\bar e^{ }_1 + \bar e^{ }_2 + 3 (\bar p^{ }_1 + \bar p^{ }_2) 
 ]/( \bar e^{ }_1 + \bar e^{ }_2 )
$.
Equation~\nr{sigw_2} implies that 
\ba
 S^{ }_{ij} 
 &
 \overset{(\bar{\varphi}\hspace*{0.3mm}')^2_{ } \,\to\, 0}{
 \underset{n^{ }_{\rm tot} \,=\, 2}{=}}
 &
 \biggl\{
 3 \, \frac{ 
 ( \bar e^{ }_1 + \bar p^{ }_1 )  \R^{ }_{1v,i}\R^{ }_{1v,j}
 + 
 ( \bar e^{ }_2 + \bar p^{ }_2 )  \R^{ }_{2v,i}\R^{ }_{2v,j}
 }{ \bar e^{ }_1 + \bar e^{ }_2 }
 \nn[3mm]
 & & \; 
 -\,2 \psi^{ }_{,i} 
 \biggl( \frac{\psi^{\hspace*{0.3mm}\prime}_{,j}}{\H} + \psi^{ }_{,j} \biggr)
 -2 \psi^{ }_{,j} 
 \biggl( \frac{\psi^{\hspace*{0.3mm}\prime}_{,i}}{\H} + \psi^{ }_{,i} \biggr)
 + \frac{2\H' \psi^{ }_{,i} \psi^{ }_{,j} }{\H^2_{ }}
 \biggr\}^\tensor_{ }
 \;, \hspace*{5mm} \la{lit_2_pre} 
\ea
whereas \eq\nr{evo_psi} can be written as 
\be
 \frac{\psi^{\hspace*{0.3mm}\prime}_{ }}{\H} + \psi 
 \;
 \overset{(\bar{\varphi}\hspace*{0.3mm}')^2_{ } \,\to\, 0}{
 \underset{n^{ }_{\rm tot} \,=\, 2}{=}}
 \;
 -\,\frac{3 
 \bigl[\, 
  (\bar e^{ }_1 + \bar p^{ }_1 )( \R^{ }_{1v}  +  \psi )
 + 
  (\bar e^{ }_2 + \bar p^{ }_2 )( \R^{ }_{2v}  +  \psi )
  \,\bigr]
 }{2 (\bar e^{ }_1 + \bar e^{ }_2) }
 \;. \la{psi_2_pre}
\ee
The goal is to massage the first row of \eq\nr{lit_2_pre} so 
that we can make use of \eq\nr{psi_2_pre}. 
However, in \eq\nr{lit_2_pre}, the background enthalpy densities
weigh the {\em squares} of the respective curvature perturbations, 
whereas in \eq\nr{psi_2_pre}, they weigh {\em linear} appearances. 
Therefore, the curvature perturbations can no longer be fully eliminated.

A convenient way to proceed is to first subtract the structure
that appears in \eq\nr{lit_1}, and to re-arrange the remainder. 
Tedious but straightforward algebra leads to 
\ba
 S^{ }_{ij}
 &
 \overset{(\bar{\varphi}\hspace*{0.3mm}')^2_{ } \,\to\, 0}{
 \underset{n^{ }_{\rm tot} \,=\, 2}{=}}
 & 
 \biggl\{ 
 \frac{4(\bar e^{ }_1 + \bar e^{ }_2)}
      {3( \bar e^{ }_1 + \bar e^{ }_2 + \bar p^{ }_1 + \bar p^{ }_2)}
 \biggl( \frac{\psi^{\hspace*{0.3mm}\prime}_{,i}}{\H} + \psi^{ }_{,i} \biggr)
 \biggl( \frac{\psi^{\hspace*{0.3mm}\prime}_{,j}}{\H} + \psi^{ }_{,j} \biggr)
 \; + \; 
 2 
 \, 
 \psi^{ }_{,i}\, \psi^{ }_{,j} 
 \nn[3mm]
 & & 
 \; + \,
 \frac{3(\bar e^{ }_1 + \bar p^{ }_1) (\bar e^{ }_2 + \bar p^{ }_2) }
 {(\bar e^{ }_1 + \bar e^{ }_2)
  ( \bar e^{ }_1 + \bar e^{ }_2 + \bar p^{ }_1 + \bar p^{ }_2 ) }
 \, \bigl( \R^{ }_{1v} \, - \, \R^{ }_{2v} \bigr)^{ }_{,i}
 \, \bigl( \R^{ }_{1v} \, - \, \R^{ }_{2v} \bigr)^{ }_{,j}
 \biggr\}^\tensor_{ }
 \;. \la{lit_2}
\ea
If one of the fluids is taken to be pure radiation
($\bar e^{ }_1\to \bar e^{ }_r$, $\bar p^{ }_1 \to \bar e^{ }_r/3$), 
and the other non-relativistic matter
($\bar e^{ }_2\to \bar e^{ }_m$, $\bar p^{ }_2 \to 0$), 
this agrees with \eq(A.34) of ref.~\cite{chicago}
(note from \eq\nr{R_v} that 
$\R^{ }_{1v} - \R^{ }_{2v} = \H ( \delta v^{ }_2 - \delta v^{ }_1 )$).
}

%
\paragraph{Role of the tensor part of anisotropic stress.}

The tensor part of the anisotropic
stress ($\Pi^\tensor_{ij}$) appears 
on the right-hand side of \eq\nr{sigw_2}. It plays there a natural
role, as a counterterm. Namely, when we take two scalar fluctuations at 
a same point, they can possess short-distance noise
(of quantum or thermal origin), and yet combine to 
give a long-wavelength signal. In momentum space, this means that
$\vec{p} + (\vec{k - p}) = \vec{k}$ even if $|\vec{p}| \gg |\vec{k}|$.
However, the physics of large momenta cannot be described by classical
field theory, and a counterterm is needed for cancelling the contribution
from the short-distance domain, whose proper treatment would require
a quantum-field-theoretic computation. 

%
\section{Summary and outlook}
\la{se:concl}

Scalar-induced gravitational waves (SIGW) have 
a partly controversial history, 
particularly in view of their gauge dependence. 
The purpose of this paper has been to demonstrate, 
for a physically relevant though, to our knowledge, 
previously unexplored 
case, how a well-defined second-order 
source term and wave equation can be derived, without assumptions about 
the dynamics of scalar perturbations, or 
the equation of state during or after SIGW generation. 

Our main result is the gauge-independent
evolution equation~\nr{sigw_2}. 
On the right-hand side, 
$\R^{ }_\varphi$ and $\R^{ }_{nv}$ are gauge-invariant
curvature perturbations, whose solution is determined by a coupled set
of stochastic differential equations~\cite{dissip,fluctu}, 
with the stochasticity sourced by thermal noise that is 
inevitably present as the universe is heating up. 
The gauge-invariant
Bardeen potential, $\psi$, is related to 
$\R^{ }_\varphi$ and $\R^{ }_{n v}$ by \eq\nr{evo_psi}. 
The anisotropic stress, $\Pi^\tensor_{ij}$, 
can incorporate viscous corrections to fluid flows, 
generate hydrodynamic fluctuations, 
or act as a ``counterterm'', cancelling contributions 
from short-distance vacuum fluctuations that would require
a quantum-field-theoretic treatment.%
\rev{%
If $ (\bar{\varphi}\hspace*{0.3mm}')^2_{ } $ 
is so small that 
$\R^{ }_\varphi$ plays no role, 
and we consider a system with 
one or two ideal fluids, 
the respective known results~\cite{sigw,chicago}
can be recovered from our more general equations, 
as shown explicitly in \eqs\nr{lit_1} and \nr{lit_2}.
}

On the left-hand side of \eq\nr{sigw_2}, 
$\vartheta^\tensor_{ij}$ represents the 
gravitational-wave perturbation, and $\Q^\tensor_{ij}$ is necessary for 
guaranteeing gauge invariance at second order. 
The latter contains non-propagating fields, which can be consistently 
set to zero in a locally flat region at the {\em observation time}.
This suggests that the gravitational-wave
energy density can be measurable, though we have not
written down an explicit observable.

We end by invoking 
one fresh motivation for considering \eq\nr{sigw_2}, 
namely the SIGW spectrum generated by a Standard Model
embedded scenario, implementing smooth reheating
via a period of sphaleron-friction-induced warm inflation~\cite{sm1}. 
There are good reasons to expect an interesting signal in this case, 
given that hydrodynamic fluctuations generate an  
$\sim f_\now^3$ infrared tail to the SIGW energy-density spectrum, 
proportional to the shear viscosity~\cite{qualitative,reheat}, 
and that shear viscosity is very large in a weakly-coupled 
$\lambda\varphi^4_{ }$ theory, 
being proportional to $1/\lambda^2_{ }$~\cite{jeon}.
However, the scenario entails other
couplings as well, notably 
from the vertex between $\varphi$ and Standard Model fields, 
which may attenuate the fluctuations of~$\varphi$. 
To reliably determine
the associated fluctuation spectrum probably
requires a numerical study, in which our source
term should play an important role.  

%
\section*{Acknowledgements}

We are grateful to Philipp Klose and Alica Rogelj for 
previous collaborations, which had 
built up the basic inspiration for the present study. 
S.P.\ thanks Ruth Durrer and Stylianos Papadopoulos for many 
helpful discussions. 

\newpage

%
\appendix
\renewcommand{\thesection}{\Alph{section}}
\renewcommand{\thesubsection}{\Alph{section}.\arabic{subsection}}
\renewcommand{\theequation}{\Alph{section}.\arabic{equation}}

%
\section{Shear part of the second-order Einstein tensor}

For completeness, 
we list here the terms for the $i\neq j$ part of the Einstein
tensor from \eq\nr{Gij_t_1}, 
before substitutions of variables 
or projections to the tensor channel:
\ba
 \G^{(h^{ }_0 \times h^{ }_0) }_{ij}
 & = & 
 2 \hspace*{0.3mm} h^{ }_0 \hspace*{0.3mm} h^{ }_{0\der ij} 
 + h^{ }_{0\der i} \hspace*{0.3mm} h^{ }_{0\der j}
 \;, \la{G_h0_h0} \\[2mm]
 \G^{(h^{ }_0 \times Y) }_{ij}
 & = & 
 \; - \, 
   h^{ }_{0\der i} \hspace*{0.3mm} Y^{ }_{\der j} 
 - h^{ }_{0\der j} \hspace*{0.3mm} Y^{ }_{\der i}
 \;, \\[2mm]
 \G^{(h^{ }_0 \times h) }_{ij}
 & = & 
  h'_0 \hspace*{0.3mm} h^{ }_{\der ij}
 + 2 \hspace*{0.3mm} h^{ }_0 \hspace*{0.3mm}
  (\partial^{ }_\tau + 2 \H ) h_{\der ij}
 \;, \\[2mm]
 \G^{(h^{ }_0 \times \vartheta) }_{ij}
 & = & 
 4 \hspace*{0.3mm} [ (\H \partial^{ }_\tau + 2 \H' + \H^2_{ }) h^{ }_0] 
 \hspace*{0.3mm} \vartheta^{ }_{\der ij}
 - h'_0 \hspace*{0.3mm} \vartheta^{\hspace*{0.3mm}\prime}_{\der ij} 
 - 2 \hspace*{0.3mm} h^{ }_0 \, 
   (\partial^{ }_\tau + 2 \H ) \vartheta^{\prime}_{\der ij}
 \nn[2mm]
 & & \; + \, 
   h^{ }_{0\der k} \hspace*{0.3mm} \vartheta^{ }_{\der ijk} 
 + 2 \hspace*{0.3mm} h^{ }_{0\der kk} \hspace*{0.3mm} \vartheta^{ }_{\der ij}
 \;, \\[2mm]
 \G^{(Y \times Y) }_{ij}
 & = & 
 2 \hspace*{0.3mm} Y \hspace*{0.3mm} Y^{ }_{\der ij}
 + 3 \hspace*{0.3mm} Y^{ }_{\der i} \hspace*{0.3mm} Y^{ }_{\der j}
 \;, \\[2mm]
 \G^{(Y \times h) }_{ij}
 & = &
 Y'\hspace*{0.3mm} h^{ }_{\der ij}
 - (\partial^{ }_\tau + 2 \H)
   ( Y^{ }_{\der i} \hspace*{0.3mm} h^{ }_{\der j}
    + 
     Y^{ }_{\der j} \hspace*{0.3mm} h^{ }_{\der i} )
 \;, \\[2mm]
 \G^{(Y \times \vartheta) }_{ij}
 & = & 
 2 [ (3 \partial^{2}_\tau + 6 \H \partial^{ }_\tau - \nabla^2_{ }) Y ] 
 \hspace*{0.3mm} \vartheta^{ }_{\der ij}
 + Y' \hspace*{0.3mm} \vartheta^{\hspace*{0.3mm}\prime}_{\der ij} 
 \nn[2mm]
 & & \; - \, 
   Y^{ }_{\der k} \hspace*{0.3mm} \vartheta^{ }_{\der ijk} 
 + 2 \hspace*{0.3mm}
   ( 
     Y^{ }_{\der ik} \hspace*{0.3mm} \vartheta^{ }_{\der jk}
   + Y^{ }_{\der jk} \hspace*{0.3mm} \vartheta^{ }_{\der ik}
   - Y^{ }_{\der ij} \hspace*{0.3mm} \vartheta^{ }_{\der kk}
   - Y^{ }_{\der kk} \hspace*{0.3mm} \vartheta^{ }_{\der ij}
   )
 \;, \\[2mm]
 \G^{(h \times h) }_{ij}
 & = & 
 h^{ }_{\der kk} \hspace*{0.3mm} h^{ }_{\der ij}
 - 
 h^{ }_{\der ik} h^{ }_{\der jk}
 \;, \\[2mm]
 \G^{(h \times \vartheta) }_{ij}
 & = & 
 2 [ ( \partial^{ }_\tau + 2 \H ) h^{ }_{\der kk} ] 
 \hspace*{0.3mm} \vartheta^{ }_{\der ij}
 + [ ( \partial^{ }_\tau + 2 \H ) h^{ }_{\der k} ] 
 \hspace*{0.3mm} \vartheta^{ }_{\der ijk} 
 \nn[2mm]
 & & \; + \, 
     h^{ }_{\der ik} \hspace*{0.3mm}
            \vartheta^{\hspace*{0.2mm}\prime}_{\der jk}
   + h^{ }_{\der jk} \hspace*{0.3mm}
            \vartheta^{\hspace*{0.2mm}\prime}_{\der ik}
   - h^{ }_{\der kk} \hspace*{0.3mm}
            \vartheta^{\hspace*{0.2mm}\prime}_{\der ij}
   - h^{ }_{\der ij} \hspace*{0.3mm}
            \vartheta^{\hspace*{0.2mm}\prime}_{\der kk}
 \;, \\[2mm]
 \G^{(\vartheta \times \vartheta) }_{ij}
 & = & 
 \;-\, 
 2 [ ( \partial^{ }_\tau + 2 \H )
    \vartheta^{\hspace*{0.2mm}\prime }_{\der kk} ] 
    \hspace*{0.3mm} \vartheta^{ }_{\der ij}
 +
   \vartheta^{\hspace*{0.2mm}\prime }_{\der kk} 
   \hspace*{0.3mm} 
   \vartheta^{\hspace*{0.2mm}\prime }_{\der ij} 
 - 2 \hspace*{0.3mm}
   \vartheta^{\hspace*{0.2mm}\prime }_{\der ik} 
   \hspace*{0.3mm} 
   \vartheta^{\hspace*{0.2mm}\prime }_{\der jk} 
 \nn[2mm]
 & & 
 \; + \, 
     \vartheta^{ }_{\der ikl} \hspace*{0.3mm}
     \vartheta^{ }_{\der jkl}
   - 
     \vartheta^{ }_{\der ijk} \hspace*{0.3mm}
     \vartheta^{ }_{\der kll}
 \;.  \la{G_theta_theta} 
\ea
A simple {\tt Mathematica} script reproducing these expressions
and verifying \eq\nr{Gij_t_2} is attached to this paper
as an ancillary file. 

%

\end{document}